\documentclass[usenatbib]{mn2e}
\usepackage{apjfonts}
\usepackage{epsfig}
\usepackage{amssymb}
\usepackage{amsmath}
\usepackage{fixltx2e} 

\newcommand{\scaleup}{}

\newcommand\plotone[1]
 {\centering \leavevmode \includegraphics[width={0.99\columnwidth}]{#1}}
\newcommand{\plotside}[1]
 {\centering \leavevmode \includegraphics[width={0.95\textwidth}]{#1}}
\newcommand{\acknowledgments}{\begin{small}\section*{Acknowledgments}\end{small}}
\newcommand\altaffilmark[1]{$^{#1}$}
\newcommand\altaffiltext[1]{$^{#1}$}
\newcommand{\etal}{et al.}

\newcommand{\msun}{M_{\sun}}

\title[Feedback and Scatter in $M_{\rm BH}-\sigma$]{The Small Scatter in BH-Host Correlations \& The 
Case for Self-Regulated BH Growth}

\author[Hopkins \etal]{
\parbox[t]{\textwidth}{ 
Philip F. Hopkins\altaffilmark{1}\thanks{E-mail:phopkins@astro.berkeley.edu},
Norman Murray\altaffilmark{2,3},
\&\ Todd A.~Thompson\altaffilmark{4,5} } 
\vspace*{6pt} \\
\altaffiltext{1}{Department of Astronomy, University of California 
Berkeley, Berkeley, CA 94720} \\
\altaffiltext{2}{Canadian Institute for Theoretical Astrophysics, 
60 St.\ George Street, University of Toronto, ON M5S 3H8, Canada} \\
\altaffiltext{3}{Canada Research Chair in Astrophysics} \\
\altaffiltext{4}{Department of Astronomy, The Ohio State University, 
140 W.\ 18th Ave., Columbus, OH 43210} \\
\altaffiltext{5}{Center for Cosmology \&\ Astro-Particle Physics, 
The Ohio State University, 191 W.\ Woodruff Ave., 
Columbus, OH 43210} }

\date{Submitted to MNRAS, March 17, 2009}
\begin{document}
\maketitle
\label{firstpage}

\begin{abstract}

Supermassive black holes (BHs) obey 
tight scaling relations between their mass and their host galaxy
properties such as total stellar mass, velocity dispersion, and potential well depth. 
This has led to the development of self-regulated models for BH growth, in which 
feedback from the central BH halts its own growth upon reaching a 
critical threshold. 
However, models have also been proposed in which feedback plays no role: 
so long as a fixed fraction of the host gas supply is accreted, relations like those 
observed can be reproduced. Here, we argue that the 
{\em scatter} in the observed BH-host correlations presents a demanding constraint
on any model for these correlations, and that it favors self-regulated
models of BH growth. We show 
that the scatter in the stellar mass fraction within a radius $R$ in observed 
ellipticals and spheroids  
increases strongly at small $R$.  At fixed total stellar mass (or host velocity dispersion),
on very small scales near the BH radius of influence, 
there is an order-of-magnitude scatter in the amount of gas that must have entered 
and formed stars. In short, the BH appears to ``know more'' about the global host galaxy 
potential on large scales than the stars and gas supply on small scales. 
This is predicted in self-regulated models; however, models where there is no 
feedback 
would generically predict order-of-magnitude scatter in the BH-host correlations. 
Likewise, models in which the BH feedback in the ``bright'' mode does not regulate the growth of the 
BH itself, but sets the stellar mass of the galaxy by inducing star formation or blowing out a 
mass in gas much larger than the galaxy stellar mass, are difficult to reconcile with the 
scatter on small scales. 

\end{abstract}


\section{Introduction}
\label{sec:intro}

The tight empirical correlations between the masses of 
supermassive black holes (BHs) and the velocity dispersion, 
masses, and binding energy or potential well depth of their 
hosts demonstrates a fundamental link between the growth of 
BHs and galaxy formation 
\citep[e.g., ][]{magorrian,FM00,Gebhardt00,aller:mbh.esph,hopkins:bhfp.obs}. 
Understanding the physical origin and consequences of these correlations 
is critical for informing models
of the co-formation of BHs and bulges, as well as theories which
relate the evolution and statistics of BH formation and quasar
activity to the remnant spheroid population. 
Likewise, the interpretation
of observations tracing the buildup of spheroid populations
and associations between spheroids in
formation, mergers, and quasar hosts 
depends on understanding the evolution of the BH-host correlations.

Although the characteristic 
spatial and mass scales of the BH and tightly correlated host properties 
are wildly different, their characteristic energy and momentum scales 
are the same.  That is, a few percent of the radiated luminosity or momentum 
from the BH growth is comparable to the binding energy/momentum of the galactic 
gas supply. Motivated by this comparison, 
attempts to explain these correlations have led to the development of a 
class of self-regulating feedback models, which  
argue that the energy or momentum released 
from an accreting supermassive black hole, even if only a small fraction 
couples to the surrounding ISM, is sufficient to 
halt further accretion onto the black hole and drive away gas, 
self-regulating growth by shutting off 
the quasar and potentially quenching star formation in 
the host. 
\citep[see, e.g.,][]{ciottiostriker:cooling.flow.selfreg.1,
ciottiostriker:cooling.flow.selfreg.2,silkrees:msigma,
murray:momentum.winds,dimatteo:msigma,sazonov:radiative.feedback,
hopkins:lifetimes.letter,hopkins:qso.all,hopkins:old.age,springel:red.galaxies}.
In these models, star formation and inflows proceed rapidly 
before the final stages of BH growth, so that 
the BH grows in a relatively fixed background potential, which sets 
the critical BH accretion rate at which the BH halts its own
subsequent growth.\footnote{Even if the trigger for rapid BH growth is a galaxy merger, 
the background potential at the galactic center relaxes sufficiently quickly that it 
is effectively fixed at the growth stage of interest
\citep[see, e.g.,][]{hopkins:bhfp.theory,hopkins:cusps.mergers,
younger:minor.mergers}.} Implicit in these models, the idea that the BH 
growth is fueling in sudden, violent events (mergers) that lead to strong 
bulge and BH growth, links the formation of both (it is less clear whether or 
not such models could succeed if growth occurred more slowly, in low 
accretion rate states, where there might be little bulge formation and/or feedback). 

The gas mass blown out by the BH may be an order of magnitude larger than the 
BH mass itself, but is still a small fraction of the total galaxy mass; it is 
the mass ``left over'' from a central starburst.
This picture is in part supported by observations of quasar-driven outflows 
that find evidence for momentum and energy content comparable to that 
required by models 
\citep[see, e.g.,][]{tremonti:postsb.outflows,reuland:highz.radio.fb,arav:outflow.abundances.covering,
nesvadba:radio.gal.feedback,prochaska:qso.outflow}, but 
remains highly uncertain. 

In contrast, 
a separate class of models have been proposed to explain the BH-host correlations
(or a subset)
without feedback or self-regulation. 
An example is this: because the observed correlation between BH mass and 
host bulge stellar mass is approximately linear ($M_{\rm BH}\propto M_{\rm bulge}$), 
if a fixed average fraction ($\sim 10^{-3}$) of the galactic baryon supply 
is able to reach the center and fuel the BH, this particular observed correlation 
can be explained. 
The efficiency of angular momentum transport
as well as the competition between gas accretion and 
star formation, explains a small mean constant of proportionality
\citep{burkertsilk:msigma,
escala:nuclear.gas.transport.to.msigma,li:nofeedback.msigma}. 
Alternatively, stellar capture rates by the central BH based on the 
average central stellar density profile also lead to an expectation
for the average BH accretion rate
\citep[see, e.g.,][]{miraldaescude:star.capture.msigma.model}. 

Although it is possible for these models to explain the mean correlation between BH mass 
and host mass, a more stringent constraint comes from the 
surprisingly small observed scatter 
in that correlation, which observations suggest is at most 
$\sim0.25-0.3$\,dex \citep[in terms of the constrained intrinsic scatter in $M_{\rm BH}$ versus 
$M_{\rm bulge}$, $\sigma$, or host potential depth/binding energy; see][]{tremaine:msigma,
marconihunt,haringrix,graham:sersic,hopkins:bhfp.obs,aller:mbh.esph}. 
It has already been argued that this scatter (compared to that in other 
correlations) suggests that it must be galaxy properties, rather than e.g.\ the properties of 
the larger dark matter halo, that are important for whatever sets the BH mass 
\citep{wyithe:msigma.scatter.vs.mhalo}. In addition, the small scatter 
puts strong constraints on models 
where the mean relation evolves 
very strongly with redshift \citep{robertson:msigma.evolution}, although 
more moderate evolution as has been suggested from some 
observations is more easily reconciled with 
the low-redshift observations \citep[see e.g.][]{peng:magorrian.evolution,hopkins:msigma.limit,
salviander:midz.msigma.evol,woo06:lowz.msigma.evolution}.

But the observed scatter also places strong constraints on whether or not 
some non-feedback mechanisms could be responsible for the observed correlations. 
It is not hard to imagine, after all, that --- on average --- 
$\sim10^{-3}$ of the galactic gas supply in most galaxies loses sufficient 
angular momentum to reach the BH radius of influence and ultimately 
accrete. What is hard to imagine is that it would {\em always} be this 
same fraction to within better than a factor $\sim2$, regardless of the huge diversity 
in observed galaxy gas fractions, gas and stellar mass density profiles, 
and kinematic states.  On large scales where viscous forces are sub-dominant 
and gas dynamics are largely gravitational, simulations of a galaxy merger 
or interaction suggest that even small changes to orbital parameters (say, 
changing relative disk inclinations, which are expected to 
vary widely, by a couple tens of degrees) 
lead to very large (nearly order-of-magnitude) changes in the amount of 
gas that loses such a large fraction of the angular momentum; from the 
galactic perspective, $\sim10^{-3}$ and $\sim10^{-2}$ are both small, gravitationally 
negligible gas masses, so why should one be ``picked out'' so narrowly?

At the opposite extreme, some models have proposed that feedback is so strong 
that the BH directly 
determines the stellar mass of the galaxy on a short timescale, 
as opposed to the feedback models already mentioned
in which the BH feedback primarily regulates its own growth.
Such a scenario implicitly
requires that the BH form in the background of a gas-dominated galaxy. 
The BH feedback 
is then assumed to either blow out all but the ``desired'' mass, which will become 
the final stellar mass appropriate for the observed correlations,
\footnote{This is 
distinct from more common models of ``radio'' or ``quiescent''-mode AGN feedback, 
in which AGN feedback regulates the final stellar mass of the galaxy over long 
timescales by heating gas in a massive group or cluster halo 
\citep[see e.g.][]{croton:sam}.} 
or to directly induce star formation, causing the formation of the appropriate mass in 
stars  \citep[see e.g.][]{king:msigma.superfb.1,king:msigma.superfb.2,
granato:sam,silk:msigma.superfb.sb,begelman:msigma.feedback.model}. In these models, then, 
the BH mass is not a function of galaxy mass; rather, galaxy mass is a function of 
BH mass. As such, the scatter in such an inverse correlation is a 
strong constraint on these models.
Moreover, because the galaxy mass 
formed at some distance from the BH depends explicitly on the BH mass, 
the simple expectation is that galaxies of a given BH mass should have 
similar stellar mass profiles near the BH where the galaxy can most directly 
``feel'' the BH mass, but with increasing scatter at larger radii, where various 
effects will introduce scatter in ``how much'' of the BH feedback successfully affects 
gas at that initial radius. 

In this paper, we use observed galaxy stellar mass profiles and observations 
of gas masses, together with simulations of gas inflows in galaxy interactions and 
mergers, to constrain the origin of the $M_{\rm BH}$-host galaxy correlations. We show 
that both observed galaxies and simulations exhibit a large scatter in the amount
of gas that loses sufficient angular momentum to reach very small radii.  Despite
this order-of-magnitude scatter, the BH preserves (or sets)  
its mass such that it is correlated with the host on much larger scales
to better than a factor of $\sim2$.
The fact that the total stellar mass on small scales has much larger scatter 
than on large scales is hard to reconcile with most 
existing  models in which the BH is simply sensitive to the {\em local} mass 
supply.  Indeed, this suggests that some process such as feedback should be present 
in order for the BH to be sensitive to {\em global} quantities such as the 
total mass of the galaxy (or total binding energy/central gas potential). 
The systematics in the radial distribution of scatter in the stellar mass
also present a strong constraint on the most extreme feedback models, in which 
galaxy mass is set by BH mass, because these models must 
explain why the stellar mass at small radii is a less accurate tracer
of the BH mass than at large radius.

\section{Observations: The Scatter in the Enclosed Stellar Mass Fraction in 
Elliptical Galaxies}
\label{sec:obs}

If the central BH mass were set by a fixed small fraction of the 
host mass reaching the central region, or if BH growth was determined 
by the central gas supply, we should expect that gas supply to 
be similarly well-correlated with the galaxy mass as is the BH mass. 
Although it is not possible to directly measure the total gas content that 
reached a given radius during BH growth,  we can estimate this quantity and set a 
strict lower limit on it by measuring the {\em stellar} mass inside 
a given radius in local spheroids. To lowest order, this will simply trace 
the total gas content that reached that radial scale. 
We discuss possible corrections to this below, but note that, strictly 
this sets a lower limit to the gas content that 
has entered a given galactic radius.

\begin{figure*}
    \centering
    \scaleup
    \plotside{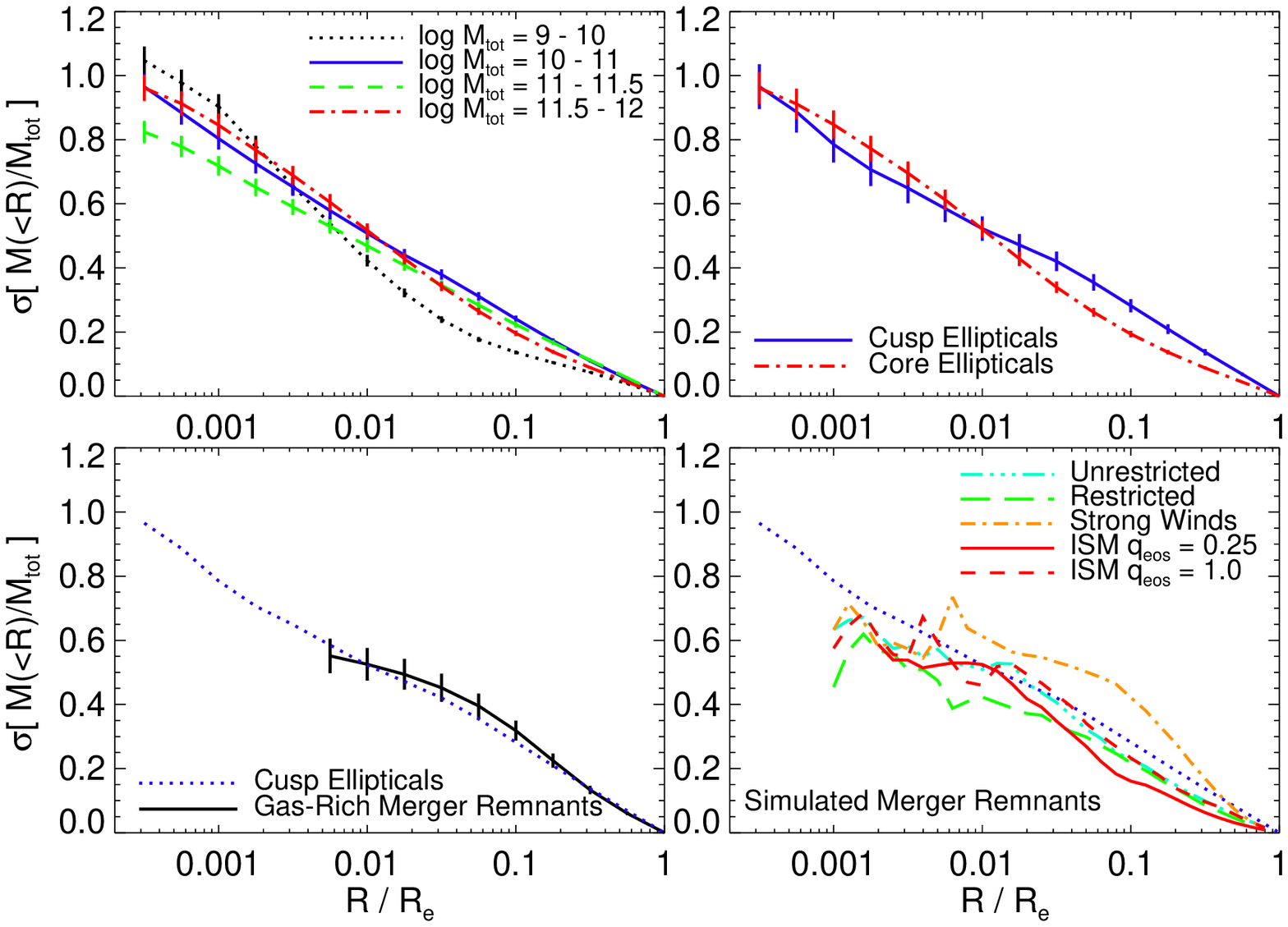}
    \caption{{\em Top Left:} Scatter in the enclosed stellar mass fraction  
    versus $R/R_{e}$ for ellipticals 
    from the samples of \citet{jk:profiles} and \citet{lauer:bimodal.profiles}, 
    in bins of stellar mass. Note $\sigma[M(<R)/M_{\rm tot}]=0$ 
    at $R=R_{e}$, by definition.
    At small radii, the scatter increases significantly.
    {\em Top Right:} Same, but for the most well-studied mass range  
    ($\sim L_{\ast}$), and with the sample divided into nuclear ``cusp'' and ``core'' 
    galaxies.  Although the median central profile shapes are different,
    $\sigma[M(<R)/M_{\rm tot}]$ is relatively independent of cusp/core status, 
    suggesting no more than weak dependence of $\sigma$ on subsequent evolution 
    and ``scouring.'' 
    {\em Bottom Left:} The ``cusp'' result compared with 
    mass profiles of observed recent gas-rich merger remnants 
    \citep{rj:profiles}. 
    The profiles in scatter agree, 
    suggesting that the diversity in stellar mass at small scales  
    is put in place by the recent star formation in gas inflows.
    {\em Bottom Right:} Comparison between the ``cusp'' ellipticals and the relaxed remnants 
    of a diverse set of {\it simulated} gas-rich mergers (see \S\ref{sec:theory}). 
    In all cases, $\sigma[M(<R)/M_{\rm tot}]$ increases at small radii.
    \label{fig:scatter.vs.r}}
\end{figure*}

Given a galaxy radial stellar surface density profile of the form $\Sigma(R)$, we 
integrate to determine the mass within $R$, $M(<R)$.
Since we are interested in comparing with BH mass,  
which follows a linear relationship with total galaxy stellar mass 
\citep[$M_{\rm BH}/M_{\rm tot}\equiv\mu_{\rm BH}\approx 0.0012$\, is assumed for 
all galaxies in the sample; see][]{haringrix}, we 
divide out the total mass and obtain the quantity of interest, 
the mass fraction within $R$, 
$f(<R)\equiv M(<R)/M_{\rm tot} \approx L(<R)/L_{\rm tot}$ (the last 
equality being for 
the observed surface brightness profile in any band where 
the mass-to-light ratio is a weak function of radius). Specifically 
\begin{equation}
f(<R)=\frac{M(<R)}{M_{\rm tot}}\equiv 
\frac{\int_{0}^{R}\Sigma(R^{\prime})\,2\pi\,R^{\prime}{\rm d}R^{\prime}}
{\int_{0}^{\infty}\Sigma(R^{\prime})\,2\pi\,R^{\prime}{\rm d}R^{\prime}}\ .
\end{equation}

The quantity $f(<R)$ is at fixed {\em physical} size: systems with 
different effective radii will have significantly different 
$f(<R)$ at the same $R$. Because the effective radius $R_{e}$ varies 
significantly as a 
function of total stellar mass, this would alone imply (even if galaxy profile 
shapes were identical) a very large variation in a population of galaxies 
in $f(<R)$ at small $R$. Even at fixed stellar mass, the factor 
$\sim2-3$ scatter in $R_{e}(M_{\rm tot})$ implies 
more than an order-of-magnitude scatter in $f(<R)$ at small $R$
if all galaxies have identical $r^{1/4}$-law \citet{devaucouleurs} profiles. 
Moreover, if the mass of the BH is
a constant fraction of the mass of the host, 
then the BH radius of influence  --- the radius that should matter for the 
gas supply available for accretion --- is a constant fraction of 
$R_{e}$ (since this radius is where the potential contributed by 
the BH and galaxy stellar mass are similar, i.e.\ 
$G\,M_{\rm BH}/R_{\rm BH} \approx G\,M_{\rm tot}/R_{e}$, hence 
$R_{\rm BH} \approx \mu_{\rm BH}\,R_{e}$). 
It is therefore more appropriate (and allows us to compare all objects 
on equal footing) to consider the mass fraction within 
a fraction of the effective radius, $f(<R/R_{e})$ (equivalently, to 
refer to all radii in units of the effective radius of the galaxy, 
rather than fixed physical units).\footnote{
By definition the mass fraction inside $R_{e}$, $f(<R/R_{e}=1)=1/2$.} 
In short, for any fraction of the effective radius (or 
multiple of the BH radius of influence), we determine the 
fraction of the galactic gas supply that was available to turn into 
stars $f(<R/R_{e})$ by measuring the stellar mass fraction inside 
that $R/R_{e}$.

The quantity of interest is, however, not the median value of $f(<R/R_{e})$ 
(that is only a restatement of the typical profile shape of ellipticals), 
but how much {\em scatter} there is in $f(<R/R_{e})$ at
$R/R_{e}$. Specifically, we consider a subsample of galaxies 
(usually at fixed stellar mass, to further marginalize over possible 
systematic differences, although we find this makes little difference), and 
determine $f(<R/R_{e})$ by integrating the surface brightness measurements 
available, from the minimum resolution elements out to $R_{e}$.
For all the objects we consider, measurements extend to radii very much larger 
than $R_{e}$. 

We then 
determine the scatter in $\log{[f(<R/R_{e})]}$, $\sigma[ M(<R)/M_{\rm tot} ]$, 
in that sample at each 
$R/R_{e}$, either by assuming the distribution is lognormal and fitting it as such, 
or by taking the IPV value to reduce bias from outliers or skewness. We find 
it makes no difference. 
In terms of the simple dispersion (rather than a more complex fitted IPV value) this is
\begin{equation}
\sigma[ f ] = \sqrt{\frac{1}{N}{\sum_{i}^{N}}(\log{f_{i}}-\langle\log{f}\rangle)^{2}} ,\ 
\label{eqn:sigma.defn}
\end{equation}
with $f\equiv f(<R/R_{e})$ evaluated for all $N$ galaxies 
at the same $R/R_{e}$, and $\langle\log{f}\rangle$ the mean 
$f$ at this $R/R_{e}$. 
This procedure effectively gives a minimum scatter in the 
stellar mass fraction that has formed at $R/R_{e}$ or $R/R_{\rm BH}$. 
Re-deriving the  results that follow in terms of absolute physical 
radii (at fixed $M_{\rm tot}$) we find the same qualitative 
results with systematically more scatter at all radii, for the reasons given above.

Figure~\ref{fig:scatter.vs.r} shows the results. 
We begin with a large sample of observed elliptical/spheroid 
surface brightness profiles from \citet{jk:profiles} and 
\citet{lauer:bimodal.profiles}, $\sim180$ unique local 
ellipticals with nuclear {\em HST} observations and 
ground-based data at large radii 
(allowing accurate surface brightness profile measurements from 
$\sim 10$\,pc to $\sim50$\,kpc). Conversion to stellar 
mass profiles and comparison of profiles from different 
instruments and wavebands are discussed extensively 
in \citet{hopkins:cusps.mergers,hopkins:cusps.ell,
hopkins:cores,hopkins:cusps.fp,hopkins:cusps.evol}; for 
our purposes the results are identical. 
We consider 
the scatter in mass fraction enclosed $\sigma[ M(<R)/M_{\rm tot} ]$ 
as a function of $R/R_{e}$ in four bins of stellar mass, where 
there are sufficient numbers of observed objects (at least 
$\sim40$ per bin) to obtain robust constraints. 

There is a clear mass-independent trend of scatter with $R/R_{e}$, 
which we find can be reasonably well approximated as 
\begin{equation}
\sigma[f(<R/R_{e})] \approx -0.28\,\log{(R/R_{e})}. 
\label{eqn:scatter.v.r}
\end{equation}
The scatter must go to zero, by definition, at $R=R_{e}$ because, by 
definition, all galaxies have exactly half their mass inside this radius, but 
the rise towards smaller radii reflects a physical diversity of profile 
shapes. Repeating this exercise in terms of physical radii in small mass 
bins, a nearly identical trend is recovered, with a systematically higher scatter 
at all radii by $\sim0.2$\,dex, and asymptoting to a constant $\sim0.2$\,dex 
scatter at radii corresponding to the typical $R_{e}$ at that mass and 
larger. 

Moreover, we have calibrated the expected results based on 
some simple experiments. For example, consider the case where we begin with a 
constant density profile over a number of 
fixed bins in $R$ (analogous to the observed points) for some number of 
test cases, then independently perturb $\Sigma(R)$ with a lognormal fixed 
scatter (say $\sim0.3\,$dex) at each $R$, and using the new density profiles 
for each system reconstruct Figure~\ref{fig:scatter.vs.r}. We 
recover the ``appropriate'' answer -- namely, that while the scatter does 
go to zero as $R\rightarrow R_{e}$, the scatter at $R\ll R_{e}$ 
converges to the input lognormal scatter as the number of mock 
systems is increased. 

Figure~\ref{fig:scatter.vs.r} shows that
at $R\sim0.1\,R_{e}$, there is a factor $2$ ($0.3$\,dex) scatter 
in the interior stellar mass fraction, and that this grows 
to a factor $\sim4$ ($0.6\,$dex) at $R\sim0.01\,R_{e}$ and 
factor $\sim8$ ($0.8\,$dex) scatter near the BH radius of 
influence, $R\sim0.001\,R_{e}$. In other words, by the time one is inside just 
$10\%$ of $R_{e}$, there is already more variation (galaxy-to-galaxy at fixed 
stellar mass) in the amount of mass that has ``made it in'' to smaller $R$ 
than there is scatter in BH masses; this difference only grows 
more and more pronounced as one moves to smaller radii. 

It is perhaps possible that some of this diversity on small scales
owes to evolution subsequent to the 
formation and growth of the BH itself. For example, 
the difference between ellipticals with central ``cusps'' and those with 
nuclear ``cores'' (steep versus flat central profile slopes, respectively) 
is commonly attributed to ``scouring,'' or the scattering 
of stars from an initially cuspy profile 
in three-body interactions by a binary BH after a ``dry'' or 
dissipationless galaxy merger \citep{begelman:scouring}. 
To check this, we repeat our comparison in 
Figure~\ref{fig:scatter.vs.r} (upper right panel) separately for galaxies classified as 
``cusp'' and ``core'' systems (classifications for 
each object are given along with surface brightness profiles). 
To obtain the most robust statistics, we consider 
a bin in total stellar mass near $\sim L_{\ast}$ ($M_{\rm tot}=10^{10}-10^{11}\,\msun$), 
where our sample includes $\sim30$ each of the cusp and core populations. 
At higher and lower masses, the comparison is similar, but the number of 
cusp and core ellipticals, respectively, drops rapidly.   
We find that there is no dramatic 
difference between the scatter within each of the two populations:
the scatter in the amount of enclosed stellar mass increases dramatically
at smaller radii for both cusp and core ellipticals.

Alternatively, we can compare the results for galaxies that are known to 
be recent gas-rich merger remnants. Specifically, we repeat our analysis 
for the sample of confirmed gas-rich merger remnants with 
near-infrared surface brightness profiles from \citet{rj:profiles}, 
also discussed in the same context as the elliptical profiles 
above in \citet{hopkins:cusps.mergers}. These objects are relatively young, 
so have not had much opportunity to be affected by subsequent evolution,
although they are sufficiently evolved such that they are dynamically 
relaxed at $R\lesssim R_{e}$. As discussed in \citet{hopkins:cusps.mergers} 
we exclude all objects with obviously unrelaxed features such as 
e.g.\ shells or tidal tails inside these radii, and check that the dynamical 
times at the radii of interest are shorter than the mean stellar population/secondary 
burst population ages. Although the resolution for these objects does not reach the 
extremely small radii of the Virgo elliptical {\em HST} sample, the sample 
still overlaps over more than two orders of magnitude in $R/R_e$.  The
lower left panel of Figure~\ref{fig:scatter.vs.r} shows that the
scatter derived from this sample agrees well with 
the trend seen in cusp ellipticals of the same mass. 
We have experimented both with the entire merger remnant sample 
taken together and that sample binned by mass, and we find that 
our conclusions and the quantitative and qualitative shape of the
scatter are unchanged.

We emphasize that our discussion of the scatter in the enclosed stellar
mass with respect to the $M_{\rm BH}-$host correlations implicitly assumes
that all galaxies in the sample lie on the observationally established 
$M_{\rm BH}-$host correlations, even though for many systems a BH mass
has yet to be established. However we have examined the $\sim10$ 
cases in the adopted samples that do have directly determined 
BH masses \citep[compiled from][]{tremaine:msigma,marconihunt,haringrix,
aller:mbh.esph} and find that 
they are consistent with the above conclusions. The observed 
BH mass and radius of influence (calculated from the observed BH 
mass and velocity dispersion profile) agree with the simple esti- 
mates based on a constant BH to host mass ratio above. The scatter 
in e.g. the ratio $M_{\rm BH}/M_{\rm gal}(<R)$ exceeds an order of magnitude for 
$R\ll R_{e}$, as predicted in Figure 1. For $R\gtrsim R_{e}$ ($M(<R)\rightarrow M_{\rm tot}$), 
on the other hand, the scatter in $M_{\rm BH}/M_{\rm gal}(<R)$ 
approaches the canonical $\sim0.3$\,dex. Bear in mind, however, that the number of 
such systems is small -- rigorously re-fitting a BH-host mass 
relation as a function of mass in different radii and estimating the internal scatter 
will require larger overlapping samples with directly determined BH masses.

\begin{figure}
    \centering
    \scaleup
    \plotone{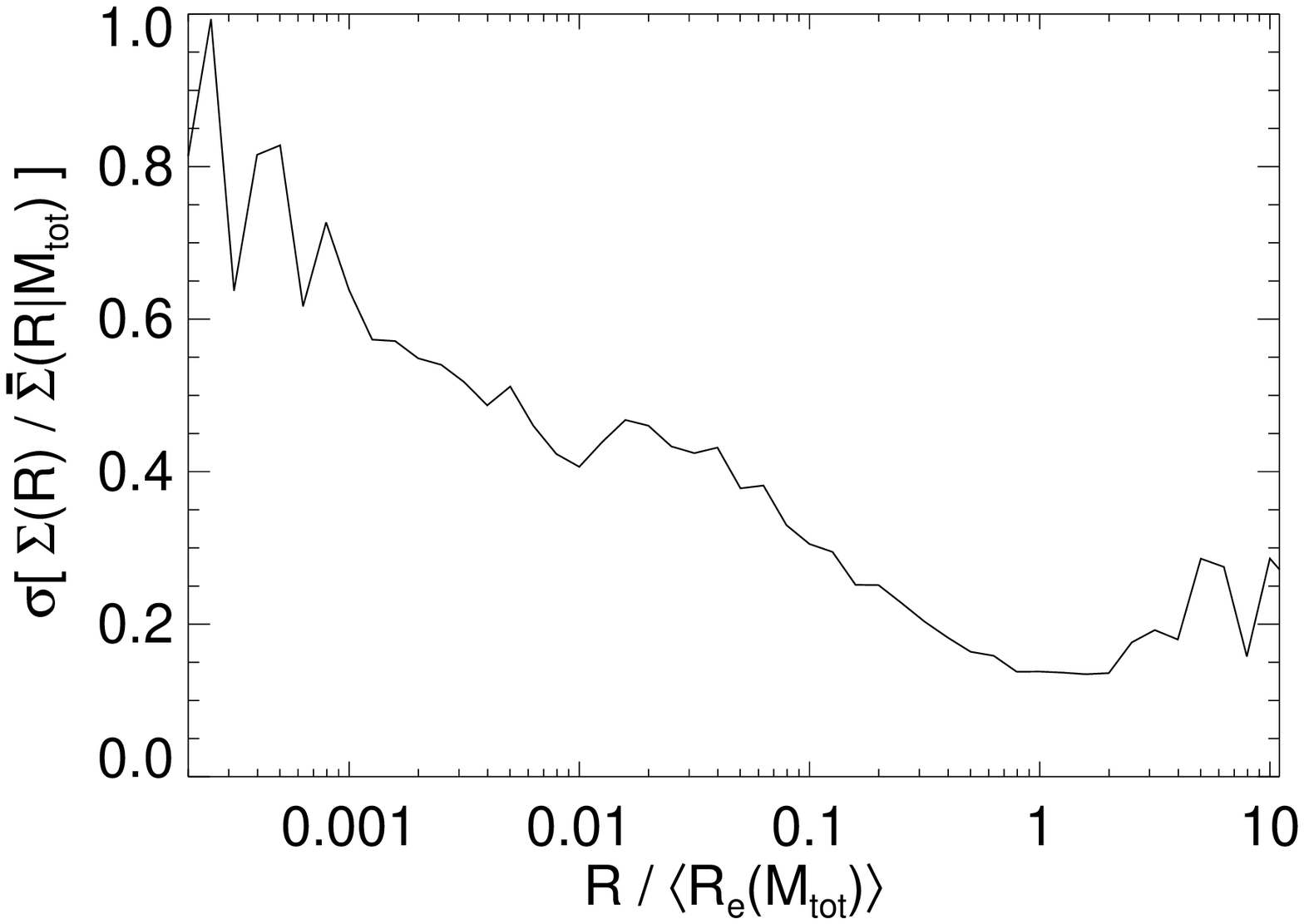}
    \caption{As Figure~\ref{fig:scatter.vs.r}, but showing the logarithmic scatter 
    in the stellar surface mass density $\Sigma(R)$ (equivalently, mass in some 
    {\em annulus} at $R$) relative to some median profile shape for ellipticals 
    of similar stellar mass ($\bar{\Sigma}(R\,|\,M_{\rm tot})$), as a function of 
    radius $R$ relative to the effective radius of that median profile
    $\langle R_{e}(M_{\rm tot})\rangle$ (the median $R_{e}$ of spheroids of the 
    given mass). In a narrow stellar mass bin, this is equivalent to 
    the scatter in $\Sigma(R)$ at each $R$. 
    Dividing the sample as 
    in Figure~\ref{fig:scatter.vs.r} yields similar results. 
    The scatter does not vanish at $\langle R_{e}(M_{\rm tot})\rangle$, 
    but it small -- this reflects the scatter in the size-mass relation. 
    The rise in scatter to small radius is similar to Figure~\ref{fig:scatter.vs.r} -- 
    the trends reflect genuine increased scatter in central properties, 
    not an artifact of the fitting/quantities plotted. 
    \label{fig:msig.tightness.new}}
\end{figure}

As noted above, the definition adopted implicitly gaurantees 
the scatter vanishes at $R=R_{e}$. In the interests of looking for a location where 
the scatter is ``minimized,'' it is useful to define a similar quantity without this 
feature (even if the 
interpretation is somewhat less intuitive). 
For example, the scatter in surface stellar mass density relative to 
some median profile of galaxies at the same mass. We show this in 
Figure~\ref{fig:msig.tightness.new}. 

At a given stellar mass, we define the ``median'' profile as 
a \citet{devaucouleurs} $r^{1/4}$ law with the 
same total stellar mass, but an effective radius equal to the median effective 
radius $\langle R_{e} \rangle$ of galaxies at that mass 
(from a quadratic fit to the $R_{e}-M_{\rm tot}$ relation in the sample). 
Knowing this $\langle R_{e}(M_{\rm tot}) \rangle$, 
the median profile is then
\begin{equation}
\bar{\Sigma}(R\,|\,M_{\rm tot}) \equiv B_{4}\,\frac{M_{\rm tot}}{\langle R_{e}(M_{\rm tot})\rangle^{2}}\,
\exp{{\Bigl\{}-b_{4}\,{\Bigl(}\frac{R}{\langle R_{e}(M_{\rm tot}) \rangle}{\Bigr)}^{1/4} {\Bigr\}}}
\end{equation}
where $B_{4}$ and $b_{4}$ are the appropriate normalization constants. 
We can then consider the ratio of the actual stellar mass surface 
density\footnote{We determine the stellar surface mass density profile 
for each object from its surface brightness profile by 
assuming a constant stellar mass-to-light ratio as a function of radius 
and normalizing to the total stellar mass. The stellar masses 
are determined from the multi-band photometry (integrated 
until the light is converged), using the color-dependent 
mass-to-light ratio calibrations in \citet{bell:mfs}, 
assuming a ``diet'' \citet{salpeter:imf} IMF. Details are 
discussed in \citet{hopkins:cusps.ell}; 
changing the IMF, bands used, or allowing for a radius-dependent 
stellar mass-to-light ratio (based on the observed color 
gradients) makes little difference.}
$\Sigma(R)$ to $\bar{\Sigma}(R\,|\,M_{\rm tot})$. 
Figure~\ref{fig:msig.tightness.new} plots the logarithmic scatter in this 
ratio at fixed value of $R / \langle R_{e}(M_{\rm tot}) \rangle$ (i.e.\ fixed 
$R$ relative to the median $R_{e}$ at a given mass): i.e.\ 
Equation~\ref{eqn:sigma.defn} where 
$f=\Sigma(R)/\bar{\Sigma}(R\,|\,M_{\rm tot})$ 
in some narrow interval (here $0.1\,$dex intervals) in 
$R / \langle R_{e}(M_{\rm tot}) \rangle$. 
Because we are interested in the scatter, the median profile 
shape is implicitly normalized out; the results in Figure~\ref{fig:msig.tightness.new} 
are nearly unchanged if we assume a different Sersic profile 
with e.g.\ $n=2-8$, or construct a non-parametric mean profile. 

The results are similar to Figure~\ref{fig:scatter.vs.r}: the scatter in surface density 
at a given radius (clearly related to the scatter in enclosed mass inside 
some radius) scales in a similar manner, reaching an order of magnitude 
at sufficiently small radii. Unlike Figure~\ref{fig:scatter.vs.r}, the scatter does 
not have to go to zero at $R=R_{e}$; rather, at $R=\langle R_{e}(M_{\rm tot})\rangle$, 
the scatter reflects that in the size-mass relation of ellipticals 
\citep[see e.g.][]{shen:size.mass}. Nevertheless, the scatter is much smaller 
around $R_{e}$ (and may even reach a minimum at these radii -- rising 
again at $R\gg R_{e}$, reflecting the diversity in outer profile shapes where 
e.g.\ the presence or absence of extended envelopes is important). Likewise, 
the conclusions are the same as Figure~\ref{fig:scatter.vs.r} if we split the sample 
by stellar mass or cusp/core status.

\section{Theory: The Scatter in the Enclosed Stellar Mass Fraction 
in Simulations}
\label{sec:theory}

We can gain some insight into how this scatter arises, and what it means relative 
to the amount of gas supply that has been available to the BH, by comparing 
with the results of numerical simulations of galaxy interactions and mergers 
that drive gas inflows, and that form realistic elliptical galaxies. 
Specifically, we consider the sample of hydrodynamic simulations in 
\citet{cox:kinematics}, \citet{robertson:msigma.evolution}, and 
\citet{younger:minor.mergers} analyzed in 
detail in \citet{hopkins:cusps.mergers,hopkins:cusps.ell,hopkins:cores,
hopkins:cusps.fp,hopkins:cusps.evol,hopkins:disk.survival}. 
These amount to several hundred unique simulations, spanning a wide 
range in progenitor galaxy masses, gas fractions, orbital parameters, 
progenitor structural properties (sizes, concentrations, bulge-to-disk ratios), 
and redshift. Most of the simulations are major (mass ratios 
$\sim$1:3$-$1:1), binary encounters, 
but they also include a series of minor mergers (from mass ratios $\sim$1:20 to 
$\sim$1:3), as well as spheroid-spheroid ``re-mergers'' or ``dry mergers'' 
(i.e.\ mergers of the elliptical remnants of previous merger simulations), 
mixed-morphology (spiral-elliptical) mergers \citep[see also][]{burkert:anisotropy,
johansson:mixed.morph.mbh.sims}, 
multiple mergers, and rapid series of hierarchical mergers. 
Our results presented below are robust to these choices in the simulations. 

The numerical calculations usually include accretion and feedback from supermassive 
black holes, as well as feedback from supernovae and stellar winds. 
However, we have performed parameter studies in these 
feedback prescriptions, and find that the 
structural properties of interest here are relatively insensitive to 
these effects \citep{cox:winds.prep,hopkins:bhfp.theory,hopkins:cusps.mergers}.
These calculations broadly reproduce 
the mean and individual profile shapes of objects in the observed samples, as a
function of stellar mass, age, morphology, and galaxy type, 
as well as the fundamental plane scaling relations of spheroids. 
Since this sample reproduces the median and scatter in profile shapes, we 
have some confidence that it represents a reasonable proxy for the 
amount of gas inflow in real ellipticals.

\begin{figure*}
    \centering
    \scaleup
    \plotside{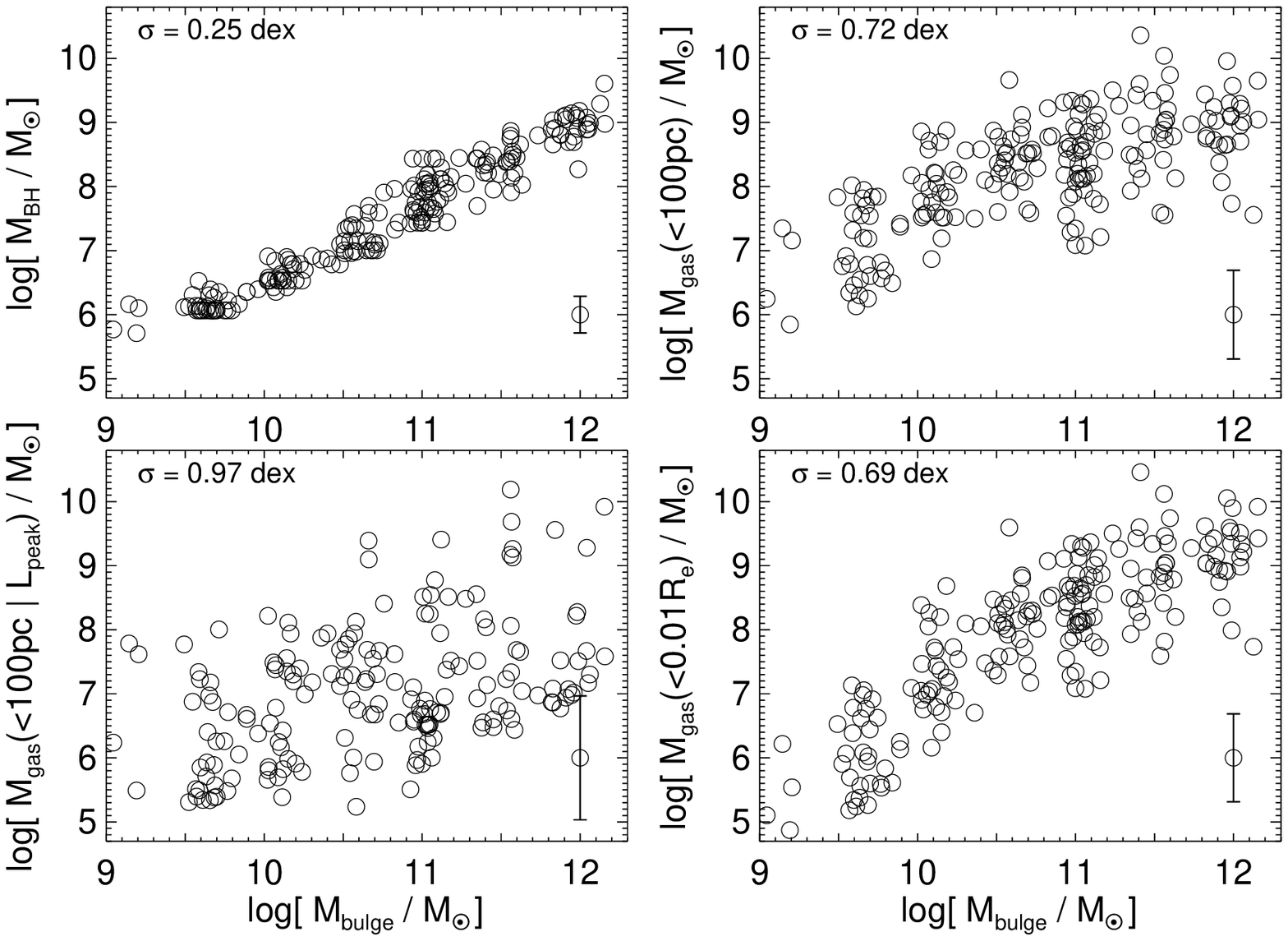}
    \caption{Results from simulated merger remnants.
    {\em Top Left:} BH mass versus host bulge mass. 
    The intrinsic scatter is shown ($0.25\,$dex). {\em Top Right:} Maximum 
    gas mass (at any time during the merger) within $100\,$pc of the BH, 
    as a function of host bulge mass. 
    A similar result is obtained if we consider the total gas mass that at some point enters the 
    central $100\,$pc, or the mass in stars in this radius, as shown in Figure~\ref{fig:scatter.vs.r}. 
    The intrinsic scatter is three times larger. {\em Bottom Left:} Same, but for the gas 
    mass inside $100\,$pc at the time when the BH is growing at its peak rate. 
    {\em Bottom Right:} Same, but for the maximum gas mass measured within a 
    fixed fraction of the remnant effective radius, rather than at fixed physical radius. 
    In these simulations, the BH mass is more sensitive to {\em global} quantities, 
    such as the stellar mass, that set the local potential depth/escape velocity, rather than 
    the local gas content. 
    \label{fig:mgas.vs.mgal}}
\end{figure*}

Figure~\ref{fig:scatter.vs.r} (bottom right panel) compares the 
scatter in interior mass fraction $\sigma[ M(<R)/M_{\rm tot} ]$ for 
this sample of simulations with the result for observed cusp ellipticals (blue
dotted line). 
We consider several sub-samples of 
simulations to see how this scatter depends on input parameters. 
First, we define an ``unrestricted'' sample that includes our entire 
suite of simulations. Second, we include a ``restricted'' sample that picks
out a subset of nearly identical simulations.  These simulations have 
galaxies with similar gas content at the time of the final merger, identical 
stellar masses, include only equal-mass mergers, and 
are chosen from a narrow range in orbital parameters. 
The scatter in this case comes, therefore, {\it only} from essentially random 
object-to-object variance in the exact dynamics of the merger, and 
consequently in the exact amount of gas that is stripped of angular momentum 
at various radii. That the scatter in the resulting stellar mass fraction
is already almost as much as that in the ``unrestricted'' sample demonstrates 
that most of the scatter in the full sample is a consequence of the 
random processes in a merger, rather than a result of the initial differences in 
gas fraction or orbital parameters.  Thus, it is difficult to marginalize the 
scatter by invoking, for example, a narrow distribution in these 
parameters at a given BH mass. 

Third, we consider simulations with very strong stellar winds; these 
simulations are discussed in detail in \citet{cox:winds.prep}. 
They include a mass loading factor several times the star formation 
rate and high wind velocities --- effectively representing the 
maximum wind feedback strength allowed by observations 
\citep[see e.g.][]{martin99:outflow.vs.m,
erb:lbg.metallicity-winds,erb:outflow.inflow.masses}. 
\citet{cox:winds.prep} demonstrate 
that yet stronger winds yield results which are physically inconsistent 
with observations: galaxies ``blow apart'' any gas concentration so 
efficiently that the existence of a starburst becomes impossible.
Thus, these ``strong winds'' simulations represent a reasonable physical 
upper limit to stellar feedback. The resulting 
dependence of scatter on radius is similar in the resulting galaxies, but with 
systematically higher normalization.

Finally, we consider a subset of simulations with two different 
specific choices for the sub-resolution prescription for the ISM 
gas \citep[specifically the ISM equation of state: effectively the pressure 
support attributed to feedback loops in the ISM such as star 
formation, supernovae, stellar winds, radiation pressure, 
cosmic rays, and other sources; see e.g.][]{mckee.ostriker:ism,springel:multiphase,
thompson:rad.pressure}. 
This is parameterized conveniently with the parameter 
$q_{\rm eos}$: $q_{\rm eos}=0$ being an isothermal equation of 
state for the ISM, the situation most unstable to e.g.\ clumping and 
gravitational fragmentation, and $q_{\rm eos}=1$ representing the 
full \citet{springel:multiphase} equation of state, which (over the 
density regime of interest) acts effectively as $P\propto \rho^{1.3-2.0}$, 
leading to an ISM more resistant to clumping and gravitational instability, 
with stronger pressure forces resisting infall and less stochastic scattering 
or torquing of gas clumps. 
Our results are similar in either case because the details 
of the gas physics or stellar feedback prescriptions are largely 
secondary with respect to gravitational torques, which dominate
the dynamics in a merger \citep{hopkins:disk.survival}.

Given these comparisons, we can see how the scatter in stellar mass content 
--- which the simulations accurately reproduce for $R/R_e\gtrsim0.001$ --- relates 
to the scatter in the gas content ``available'' to fuel the BH. 
Figure~\ref{fig:mgas.vs.mgal} shows this comparison. Specifically, we compare 
the correlation between BH mass and host spheroid stellar mass from the simulations
\citep[see][]{dimatteo:msigma,robertson:msigma.evolution},
with a small lognormal scatter of $\approx 0.25\,$dex, 
to the correlation between a number of measures of the nuclear gas and the 
bulge mass. First (upper right-hand panel), we consider the maximum gas mass at any time during the 
merger that is available inside of a nominal small radius (here $100\,$pc; the results 
scale with radius in the manner seen in Figure~\ref{fig:scatter.vs.r}; 
at smaller radii, however, our simulations begin to suffer from resolution limitations). 
We obtain a nearly identical result if we consider the {\em total} gas content 
that enters or is supplied to a given radius (and of course, the total gas content that 
ultimately stays in a given radius and is turned into stars is shown in 
Figure~\ref{fig:scatter.vs.r}). The scatter is clearly much larger than that between 
BH mass and host bulge mass, specifically in this case $0.72$\,dex. 
Second, we consider the gas mass in this radius (``available'' for accretion -- i.e.\ 
having made it to less than $100\,$pc inside the center of the starburst/gas 
inflow) at the time of the peak accretion rate/intrinsic luminosity of the BH -- 
i.e.\ just before feedback from the BH begins to regulate growth of the system. 
Here (lower left-hand panel), the scatter is even larger, almost a full order of 
magnitude ($0.97\,$dex). 
Finally, in the lower right-hand panel, 
we consider the maximum gas mass inside $1\%$ of the effective radius 
(or fixed multiple of the BH radius of influence), analogous to 
Figure~\ref{fig:scatter.vs.r}.\footnote{Here, we consider 
the true three-dimensional effective radius of the system.} The scatter is 
similar to that in the fixed $100\,$pc radius, and to that expected 
from Figure~\ref{fig:scatter.vs.r}, $0.69$\,dex. 

Regardless of how we define 
the central gas content available for BH accretion, 
it seems that the scatter in the
nuclear gas supply is always much larger than the scatter in BH mass 
at fixed host mass, in these simulations. 
Identical results are obtained in terms of other host 
properties like the velocity dispersion or binding energy. 

It is interesting to compare these predictions to alternative, 
more extreme feedback models, in which the BH mass and resulting 
feedback set the total stellar mass itself by inducing star formation 
and/or blowing out all but some desired amount of gas, which 
will then (later) form stars. This contrasts strongly with the simulations 
above, where the BH forms in a largely fixed background 
potential of stars, and self-regulates is own growth (with the mass 
blown out by the BH being a small correction to the total bulge 
mass). In the models where the BH induces the bulge formation, 
the mass formed at some $R$ is an explicit function of $M_{\rm BH}$ or BH 
luminosity: $M(< R)$ is a function of $M_{\rm BH}$, rather than the other way 
around. As such, it is difficult to explain why the central regions of 
the galaxy near the BH radius of influence show a large scatter with 
respect to $M_{\rm BH}$, while the mass formed at large radii, where the BH 
gravity is negligible, appears to reflect $M_{\rm BH}$ accurately. Quantitatively, 
we can construct analogous -- albeit 
highly simplified -- predictions for these models to 
compare with the standard self-regulated BH growth scenario simulated 
above. Specifically, we can repeat the calculations used to 
derive the expected BH-host mass relation in the models, allowing 
for some fluctuation in gas profile shapes. For example, we 
follow \citet{king:msigma.superfb.1} and assume that BH feedback drives a 
compressive shock through a large gaseous body, inducing the formation 
of a bulge on the observed relation. Specifically, we assume 
that the gas initially follows a universal profile $\bar{\rho}(r)$ 
(in this case isothermal or \citet{hernquist:profile}, but it makes little difference for 
any $\rho\propto r^{-\beta}$ with $\beta<3$, appropriate for the radii 
$r < R_{e}$ that we are considering), and that the feedback drives a cold, infinitely thin 
shell which forms stars according to the observed \citet{kennicutt98} 
relation; the details of the shell evolution are derived in 
\citet{king:msigma.superfb.1,king:msigma.superfb.2}. 
Now allow for some noise spectrum in object-to-object density profiles
$\rho(r) = \bar{\rho}(r)\,(1+\delta(r))$ ($\delta(r)$ is the amplitude of 
fluctuations in density per logarithmic interval in $r$, 
which we can parameterize
\footnote{Equivalently, if there is some characteristic size scale $r_{0}\propto r^{\eta}$ 
for fluctuations in $\rho$ and the density is a local gaussian random field, then 
$\gamma=(1-\eta)/2$. Equal magnitude of random density fluctuations/deviations per 
logarithmic interval in $r$ corresponds to $\gamma=0$ or $\eta=1$ 
($r_{0}\propto r$). Constant 
physical size-scales for characteristic fluctuations give $\gamma=1/2$. 
If the fluctuations seen in Figure~\ref{fig:scatter.vs.r} 
were put in as an initial condition in the density 
profile, it would be $\gamma=-0.3$ ($\eta=1.6$).}
as $\delta(r)\propto r^{\gamma}$) -- this can represent clumping, 
fragmentation of shells, slightly different gas thermal states, inflow rates, 
or other factors. Integrating gives a prediction for the scatter in the 
log of the enclosed mass, as a function of radius: very approximately 
$\sigma[f(<R/R_{e})] \propto \gamma + 0.5\,\log{R}$. This is straightforward to 
understand: as the blastwave propagates away from the BH, the 
``knowledge'' of the BH mass (accuracy with which the feedback 
strength at $R$ reflects the true BH mass) executes a random walk. It 
is, however, opposite the observed trend (Equation~\ref{eqn:scatter.v.r}) for most 
reasonable input noise spectra. 
This clearly represents a strong constraint on these models. But it may be 
possible to reconcile them with the observations -- for example, by invoking the 
observed trend as part of the initial conditions, imposing an 
initial spectrum of density perturbations that already resembles 
what is observed.\footnote{In detail following the calculation 
above, the initial perturbation spectrum 
would have to be tuned as a function of radius such that 
$\gamma=-0.8$ ($\eta=2.6$); 
the fractional amplitude of local density perturbations reflecting the 
location within the galaxy and characteristically about two orders of magnitude 
larger at $\sim100$\,pc than at the effective radius.} 
This may be possible if, as demonstrated in 
the numerical simulations considered, gas is torqued and can move 
gravitationally in clumps and streams through an already largely 
formed stellar/dark matter background. However recall that in these 
models the background must (by construction) be primarily a 
self-gravitating gas-dominated system. It is unclear whether or not there 
actually exist equilibrium gas density profiles that can support (in 
an object-to-object sense) variations with as steep a spectrum as 
$\gamma\sim-1$ (for the observed scatter, this would imply that the same 
median equilibrium large-scale profile could support variations of 
an order of magnitude in the density inside $\sim100$\,pc, for example). 
f not so, other radius-dependent sources of scatter would need to be 
invoked to reconcile these models and the observations. 

\section{Conclusions}
\label{sec:discuss}

We have compared the scatter in the stellar mass in the central regions of 
observed spheroids, as a proxy for the amount of gas
which was available at a given radius for possible accretion by the central
super-massive BH.  The scatter in this proxy for gas supply, from galaxy to galaxy, at 
fixed stellar mass, increases rapidly at small radii.  
It is nearly an order of magnitude around the BH radius of influence. 
We show that this is true for both observed nuclear cusp and core ellipticals, 
and for observed recent gas-rich merger remnants (see Fig.~\ref{fig:scatter.vs.r}; 
\S\ref{sec:obs}).

Despite this order-of-magnitude variation on small scales,  
the observed correlations between BH mass and 
global large-scale parameters of host spheroids (e.g., total 
stellar mass, velocity dispersion, binding energy) exhibit 
only factor $\sim2$ scatter.  In other words, the BH seems 
to ``know more'' about the host on large scales than on small scales.
This is a natural expectation in feedback-regulated models: BHs stop growing 
once they reach a sufficient mass/accretion rate to unbind material near them, 
regardless of the total fuel supply. However, this finding is 
difficult to understand in the context of 
classes of models in the literature that do not invoke feedback and 
self-regulation, but instead posit that a fixed, small fraction, of the 
total host mass is incorporated into the central BH.
If this were the case --- if this gas supply set the BH mass --- then the scatter 
in the nuclear gas supply would likely be directly related to the scatter in BH mass,
and it should decrease or stay constant towards smaller radii.  In fact, 
this is not observed.

As a check, and in order to provide context, we compare the observed scatter with 
hydrodynamic merger simulations that include  feedback from growing BHs 
(see lower right-hand panel of Fig.~\ref{fig:scatter.vs.r}, Fig.~\ref{fig:mgas.vs.mgal}, and 
\S\ref{sec:theory}).
We find that the observed trends are broadly
reproduced by the simulations. At various points in a galaxy's history,
mergers and interactions efficiently remove angular momentum from the gas, 
such that some of the gas (at some initial radii) 
effectively free-falls until shocking near the galactic center. 
As a consequence, the scatter in the amount of gas that reaches a given small radius
is set primarily by 
gravitational physics, rather than the details of the gas microphysics or feedback prescriptions.
We demonstrate explicitly that the scatter is also a general result of even very 
similar  initial conditions: at small radii, the chaotic nature of mergers and 
interactions is important, and even strong restrictions in  the initial gas 
fractions, redshifts, and orbits of interacting systems does not significantly
decrease the scatter in gas supply on scales comparable to the BH radius of influence. 
We also find that the simulations validate our assumption that the enclosed stellar
mass is a faithful proxy for the supplied gas mass.
Of course, it remains to be seen if this will be true as well for more 
``quiescent'' fueling mechanisms (e.g.\ galactic bars and minor mergers, 
or stellar mass loss), that may be important or even dominant in the 
AGN population. 

These findings relate to another interesting property of spheroids: 
the absolute value of the central/peak stellar surface density of spheroids 
is observed to be relatively constant (factor $\sim3-5$ scatter) independent of 
e.g.\ total stellar mass or effective radius of the galaxy 
\citep[see e.g.][]{lauer:bimodal.profiles,jk:profiles,hopkins:cusps.ell}, 
which vary widely. If BH mass were set purely by local processes, the 
simplest expectation from this fact would be that all galaxies should correspondingly 
have the same BH mass (and that, as a consequence, there should be 
very large scatter between BH mass and {\em total} galaxy mass or 
velocity dispersion). Instead, it seems to be these integral properties, 
rather than local density, that correlate best with BH mass. 

The constraints we derive on the run of observed scatter in stellar mass 
also appear constrain models at the opposite extreme. In these, 
feedback from the accreting BH is so strong that it determines the stellar mass of 
the galaxy, either by directly inducing star formation --- essentially triggering 
bulge formation --- or by blowing out a mass in gas much larger than the 
final stellar mass of the galaxy. In these 
models, galaxy mass is a function of BH mass, rather than the other way around. 
The scatter in the BH-host mass correlations is therefore a direct constraint 
on these models: in particular, not the absolute magnitude of the scatter, but the 
trend in scatter with radius is of interest. The stellar bulge mass formed in some 
shell at a given radius around the BH depends explicitly in these models 
on the BH mass (or luminosity), since this is what is triggering star formation or 
removing the gas supply that would otherwise make many more stars. 
Intuitively, it is difficult then to explain what is observed in Figure~\ref{fig:scatter.vs.r};
despite the fact that the mass in each annulus should 
be determined by $M_{\rm BH}$ (if this class of models were true), 
the central regions of the galaxy show more scatter with respect 
to $M_{\rm BH}$ than the outer regions of the galaxy. 
In particular, the bulge mass 
at radii close to the BH radius of influence 
appears to have relatively little knowledge of the BH mass, 
exhibiting an order-of-magnitude scatter at given $M_{\rm BH}$.  In contrast, 
the mass formed at large radii, near the effective radius of the galaxy,
here the BH's gravity is completely negligible
appears to trace the BH mass accurately.
There may be particular initial conditions or other controlling parameters 
that explain this, but it is clear that the observed scatter can strongly 
constrain such models. 

The systematics of the observed scatter in the BH-host relations therefore --- 
perhaps moreso than the normalization or slope of the relations  ---
suggests that BHs may indeed 
self-regulate at a critical mass determined by global host galaxy properties 
in the manner predicted by feedback models. At the very least, non-feedback 
models must be revisited. It is insufficient to predict the normalization and slopes 
of the BH-host correlations. They must also provide predictions that 
reproduce the small observed scatter at large radii and its
significant increase at small radii.
Likewise, models of very extreme feedback, in which the BH mass is not set 
by galaxy properties but rather, galaxy properties are set by BH mass, 
must explain how the central regions of the galaxy closest to the BH
appear relatively insensitive to the BH mass. 
There is also the constraint from the BH ``fundamental plane'' -- that 
it appears that neither $M_{\rm BH}-\sigma$ nor $M_{\rm BH}-M_{\rm bulge}$ 
is most fundamental, but rather a combination that 
traces bulge binding energy ($\sim (M_{\rm bulge}\,\sigma^{2})^{0.7}$) 
\citep[all models should predict, for example, that $M_{\rm BH}$ correlates with 
$\sigma$ at fixed $M_{\rm bulge}$ and vice versa, as demonstrated in 
the observations in][]{hopkins:bhfp.obs}. Thus far,  
the class of self-regulating feedback models appear to be 
most successful at simultaneously explaining these observations, but more work is 
necessary. 

Finally, we note that the simulations described in \S\ref{sec:theory}
indicate that the total stellar mass is a better predictor of $M_{\rm BH}$ than 
e.g.\ the stellar mass inside some small radius $R$. This prediction, 
coupled with the results in Figure~\ref{fig:scatter.vs.r}, indicate that 
future observations of the galaxies that comprise the observed 
$M_{\rm BH}-$host relations should also consider the radius-dependence of 
those relations (i.e.\ $M_{\rm bulge}-M_{\rm tot}(<R)$ or $M_{\rm BH}-\sigma(R)$ 
correlations, as a function of $R$). Comparison of such correlations, and identification 
of scales where the scatter in the relations is minimized, 
can significantly constrain models of self-regulated BH growth.

\acknowledgments 
We thank Eliot Quataert, Carlos Frenk, and Lars Hernquist for helpful discussions. 
We also appreciate the hospitality of the 
Aspen Center for Physics, where this paper was partially developed. 
Support for PFH was provided by the Miller Institute for Basic Research 
in Science, University of California Berkeley.
\\

\bibliography{/Users/phopkins/Documents/lars_galaxies/papers/ms}

\end{document}